\documentclass[preprint]{aastex}

\usepackage{graphicx}
\usepackage{layout}
\usepackage[latin1]{inputenc}
\usepackage{amssymb}
\usepackage{natbib}

\begin{document}
\title{ESO-VLT optical spectroscopy of BL Lac objects: I. new redshifts.}
\author{B. Sbarufatti\altaffilmark{1}, A. Treves}
\affil{Universit\`a dell'Insubria, Via Valleggio 11, I-22100 Como, Italy}
\author{R. Falomo}
\affil{INAF, Osservatorio Astronomico di Padova, Vicolo dell'Osservatorio 5, 
 I-35122 Padova, Italy}
\author{J. Heidt}
\affil{Lanenssterwarte Heidelberg, K\"onigstuhl, D-69117 Heidelberg, Germany}
\author{J. Kotilainen}
\affil{Tuorla Observatory, University of Turku, V\"ais\"al\"antie 20, FIN-21500
 Piikki\"o, Finland}
\author{R. Scarpa}
\affil{European Southern Observatory, 3107 Alonso de Cordova, Santiago, Chile}
\altaffiltext{1}{also at Universit\`a di Milano-Bicocca}

\begin{abstract}
We report redshift measurements for 12  BL Lacertae objects from a program 
aimed at obtaining high signal to noise (up to $\sim$ 500) optical spectroscopy
of a mixed sample of objects. 
The new observations, gathered with the 8 m ESO Very Large Telescope, allowed us
to detect weak spectral features down to  a line equivalent width as small as 
$\sim$ 1 \AA. 
The new redshifts fall in the 0.2$-$1.3 interval. 
For nine objects we observe emission lines from the active nucleus. In the 
remaining three cases absorption lines from the host galaxy are found. 
For two objects we also  detect absorption lines from intervening  systems.
\end{abstract}
\keywords{BL Lacertae objects: general}
\section{Introduction}

BL Lac objects are active galactic nuclei (AGN) exhibiting strong 
nonthermal emission, superluminal motion, rapid and large flux and 
polarization variability. 
In contrast with other classes of AGN, their spectra are characterized by the 
absence or extreme weakness of emission lines. 

In the standard model of BL Lacs, the weak emission lines (if present) are 
generated by fluorescence, as in other types of AGN. The line  equivalent width (EW)
 is 
reduced by the strong beamed continuum, caused by the alignment with our line 
of sight of the relativistic jet produced by the nucleus. 
On the other hand absorption lines  can be produced either from spectral 
features of the stellar population of the host galaxy or from intervening halos
as in the case of quasars (mainly for high redshift sources). Detectability of
the absorption features depends inversely on the brightness state of the central source.  
Because of the weakness of the emission lines (usually EW$\lesssim$5 \AA) and the 
relatively bright central continuum source with respect to the emission from 
the host galaxy, in many cases the redshift of these objects is unknown and/or 
very uncertain. 
The obvious implication of this is that the distance of the sources remains 
undetermined, hampering a proper evaluation the physical parameters of the 
objects. 

In the past decade a number of projects were carried out to derive the redshift
of BL Lac objects either for selected targets or to obtain as much redshift  
information as possible for complete samples of BL Lacs. 
Apart from \citet{heidt04} all previous works  \citep[e.g.][]{stickel93, veron93,
bade94, veron94, marcha96,drinkwater97, landt01, rector01, londish02, 
carangelo03, hook03} were based on optical spectra collected with $\leq$4 m 
class telescopes and are therefore limited by either relatively low S/N ($\lesssim$50)
or to bright objects (m$_V<$15).
As a result of this in spite of the use of relatively large aperture telescopes
in a number of cases the redshift remained unknown. Since the detection of 
spectral features critically depends on the S/N of the spectra it is clear that
a further step towards the knowledge of the redshift of BL Lacs requires the use 
of 8m class telescopes.

With these aims in mind we have carried out a project to secure optical spectra
of BL Lacs of still unknown or uncertain redshift with the highest possible S/N 
using VLT in service mode. Such a program can be executed even during non 
photometric sky and poor seeing conditions and does not interfere with 
higher priority programs performed at large aperture telescopes.
We selected sources from various lists of BL 
Lacs \citep[e.g.][]{padovani95a, veron03} with 15$<$m$_{V}<$22, and 
$\delta<$20$^{\circ}$ thus ensuring to obtain high S/N spectra with reasonable
exposure times at VLT. Under these assumptions we constructed a sample of 
$\sim$60 targets. 31 have been observed in the first two campaigns while the 
others are scheduled. 
Of these observed sources 12 have featureless spectra, 2 turn out to be 
galactic stars, 2 are early type galaxies, 3 are high redshift quasars and 12 
are BL Lacs for which the redshift has been measured by us.

In this paper we report the results concerning these 12 objects.
A full account of the observations for the featureless BL Lacs and the other 
sources will be presented in a later publication. The outline of this work is as follows: in 
section 2 we describe the 
observations and the data analysis; the results for individual objects are 
presented in section 3 and  finally we give in section 4 a brief discussion of 
our findings. 

To compute absolute quantities we adopted a cosmology with 
H$_0$=70 km s$^{-1}$ Mpc$^{-1}$, $\Omega_M$=0.3, $\Omega_{\lambda}$=0.7.

\section{Observations and data analysis}

Optical spectra  were collected in service mode in Paranal (Chile) with VLT UT1
(Antu) 
equipped with FORS1 \citep{fors} in the period April 2003 to  March 2004. 
We used the 300V+I grism combined with a 2'' slit, yielding a dispersion 
110 \AA/mm (corresponding to 2.64 \AA/pixel) and spectral resolution=15--20 \AA \ covering the 3800$-$8000 \AA \ 
range. The seeing during observations was in the range 0.5$-$2.5'', with an average of $\sim$1''.
A list of the observed objects, along with relevant informations, is given in 
table \ref{tab:list}.

Data reduction was performed using IRAF\footnote{IRAF is distributed by the 
National Optical Astronomy Observatories, which are operated by the Association
of Universities for Research in Astronomy, Inc., under cooperative agreement 
with the National Science Foundation.}\citep{tody86,tody93} following standard 
procedures for spectral analysis. 
For each target we obtained three individual spectra in order to produce 
adequate correction of cosmic rays and, eventually, to provide independent 
check of weak features. The individual frames were then combined into a single 
average spectrum and processed. 
Wavelength calibration was  performed using the spectra of a Helium Neon Argon
lamp obtained during the same observing night. This allows one to obtain 
wavelength calibrated spectra with an accuracy of $\sim$ 3 \AA (rms). From 
these  calibrated images we extracted one-dimensional spectra adopting an 
optimal extraction algorithm \citep{valdes92} to improve the S/N. 

Although this program did not require good photometric conditions most of the 
observations were obtained during clear nights. This enables us to perform a 
spectrophotometric calibration of the acquired data using observations of 
standard stars \citep{oke90} observed in the  same nights. 
From the photometric database at Paranal we estimate that a photometric 
accuracy of 10\% was reached during our observing nights. 
The spectra were  also corrected for Galactic extinction, using the extinction law by
\citet{cardelli89} and assuming values of E(B-V) from \citet{schlegel98}. 
We calculated the spectral index of the dereddened spectra, fitting to the
continuum a simple power-law $F_{\lambda}\propto\lambda^{-\alpha}$.

\begin{figure}[htbp]
  \includegraphics[scale=0.6,angle=-90]{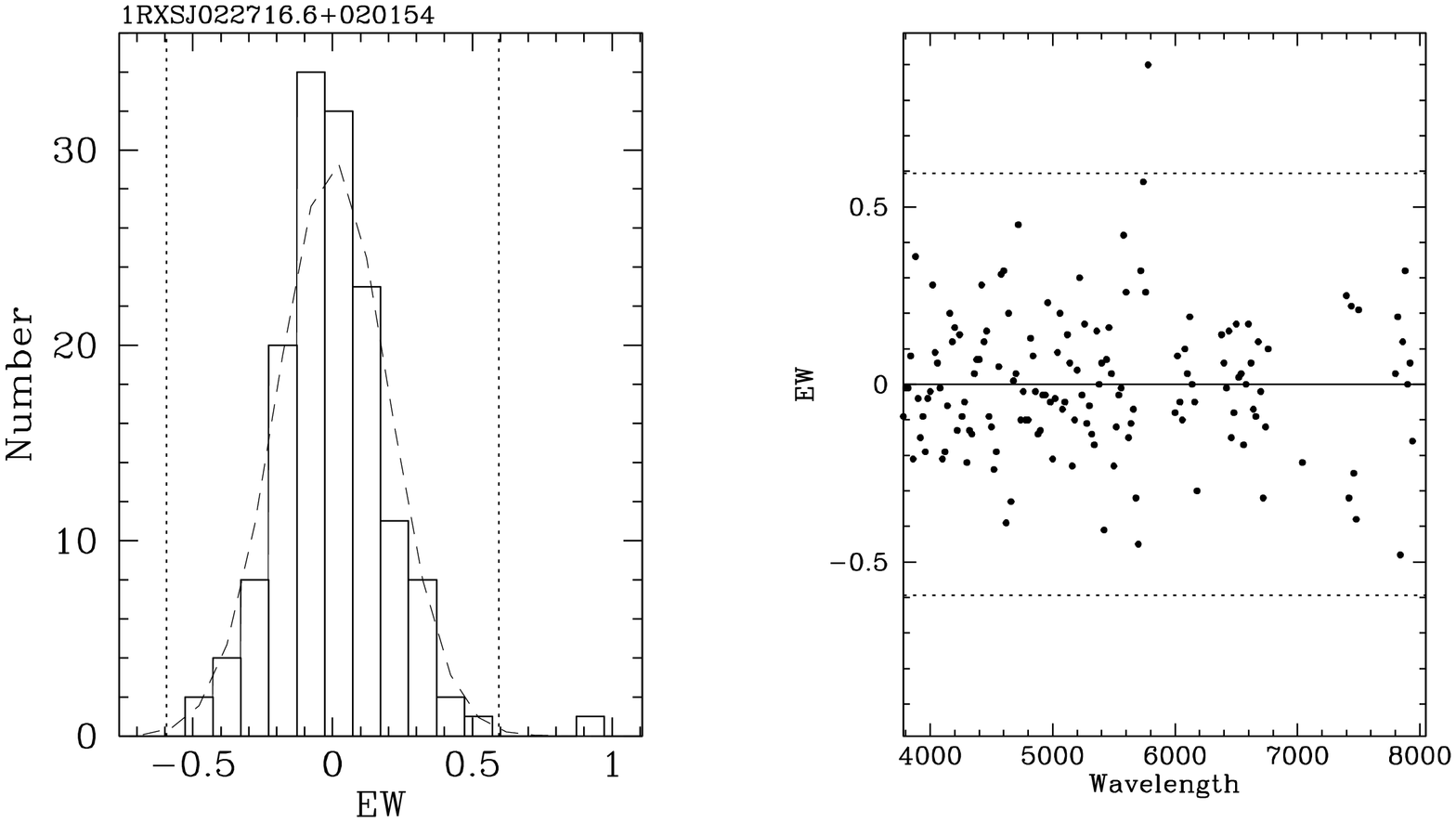}
  \caption{Distribution of EW for the object 1RXJS022716+020154.
Left panel: histogram of the distribution of the EW values. Dashed lines 
shows the gaussian fit to the distribution; dotted lines
represent EW$_{min}$. Right panel: EW distribution as
a function of the wavelength. 
Emission lines correspond to negative values of EW, absorption to 
positive values.}
  \label{fig:ewdist}
\end{figure}

In order to detect faint spectral features we evaluate for each spectrum the 
minimum measurable EW. 
This quantity was derived computing the EW for each interval of 
wavelengths of fixed size (20 \AA) along the whole spectrum but excluding the 
telluric 
absorptions. In each spectral bin we evaluate the EW, as the integral of the 
(F$_\lambda$-F$_{cont}$)/F$_{cont}$ where F$_\lambda$ is the total flux inside
the bin, and F$_{cont}$ is the continuum flux inside the bin. 
F$_{cont}$ is 
obtained from a linear interpolation of the fluxes in the adjacent bins. 
Where no spectral features are present the EW measurements yields a value 
around zero and their distribution gives a representation of
the noise in the actual spectrum. 
We assume that a spectral line is significant when its EW is larger than 3 
times the rms of the distribution. This is also taken as the
minimum detectable EW (EW$_{min}$).
An example of this procedure is shown in Figure \ref{fig:ewdist}.  
All the features above the threshold were considered as line candidates and 
were carefully visually inspected.
For the spectra presented here the EW$_{min}$ ranges from $\sim$ 1 \AA \ 
to 0.1 \AA  \ and  it is clearly well correlated with  the S/N of each 
spectrum, which was taken in a 100 \AA \ wide bin, centered at 6100 \AA \ (see 
Figure 2). 
Line centers and FWHM were determined using a gaussian fit to the line 
profile.

\section{Results for  individual sources }

The calibrated  spectra, corrected for the galactic extinction, are given in 
Figure \ref{fig:spec}. Line identifications and parameters, redshift estimates, observed V magnitudes 
(obtained from the monochromatic flux at 5500 \AA)
and continuum slopes are given in table \ref{tab:lines}. The observed 
magnitudes of our targets are consistent with the values reported in the 
\cite{veron03} catalogue within $\sim$0.5 magnitudes. This is well within the range of
the optical variability of BL Lac objects. The typical 
uncertainty of the measured redshift is $\Delta(z)\sim$0.001.
We report here some comments on individual sources:

\paragraph{1RXS J022716.6+020154} 
Previous spectra obtained by \citet{nass96} are featureless. 
Images taken by \citet{nilsson03} detected the host galaxy with
m$_{R}$=19.6. 
Our high S/N spectrum clearly 
exhibits absorption features of the host galaxy (Ca II and G band absorptions) 
at z=0.457  (corresponding to a host galaxy magnitude M$_{R}$=-23.1).
This agrees with the photometric estimate of the redshift (z$\sim$0.45) by \citet{nilsson03} 
based on the host galaxy position on the fundamental plane.
The G band 4305 \AA \ is contaminated by telluric absorption. 
Absorption features of the interstellar medium (ISM) of our 
galaxy are also detected: diffuse interstellar bands (DIB) at 4428 \AA \ and 
4726 \AA, and NaI at 5892 \AA.

\paragraph{PKS 0306+102} On the basis of a single emission line identified with
MgII 2798 \AA, \citet{veron94} proposed a tentative redshift of 0.863. Our 
spectrum clearly shows this line together with CII] 2326 \AA, [NeV] 3426\AA, 
[OII] 3727 \AA \ and [NeIII] 3869 \AA. The spectrum shows also the NaI 5892 \AA
\ absorption feature from the ISM of our galaxy. This yields a firm redshift 
determination of z=0.862. We note that, on the basis of the EW 
of the emission lines, this object seems to be of intermediate nature between a
BL Lac and a polarized QSO, as already suggested also by \citet{veron94}.

\paragraph{1RXS J031615.0$-$26074 } 
Previous spectra obtained by \citet{bade94} are featureless. 
In our spectrum we observe a faint [OII] 3727 \AA \ emission line, the
CaII 3934, 3968 \AA \  and G band 4305 \AA \  absorption features at 
z=0.443. 
The NaI 5892 \AA \ absorption from the ISM of our galaxy is also detected.

\paragraph{PKS 0338-214} \citet{wright77} reported a tentative redshift 
estimate of z=0.048, based on the detection of H$_{\beta}$, Mg H and 
H$_{\alpha}$. These features were not confirmed by \citet{falomo94} and 
\citet{falomo00}. Based on the host galaxy detection (m$_{R}$=18.86),
\citet{falomo00} proposed a photometric redshift of $\sim$0.45. Our VLT 
spectrum clearly shows the [OII] 3727 \AA \ and [OIII] 5007\AA ~  emission 
lines at z=0.223. At this redshift the host galaxy has an absolute magnitude 
M$_{R}$=-21.7. We observe also the NaI 5892 \AA \ absorption line and the 5772 
\AA \ DIB produced by our 
galaxy ISM.

\paragraph{PKS 0426-380} The only spectral feature detected by 
\citet{stickel93} is the intervening system at z=1.030. Consistently with the 
high redshift \citet{urry00} did not detect the host galaxy in their HST image.
\citet{heidt04} proposed z=1.111 based on the detection of a single emission 
line identified with the MgII 2798 \AA. 
In our spectrum we detect MgII 2798 \AA \ at z=1.112, and also CIII] and [OII] 
3727 \AA \ at z=1.098 and 1.099, respectively. There is a velocity difference 
of $\sim$1800 km s$^{-1}$ between MgII and the other lines.  This is not 
uncommon for AGN \citep[e.g.][and references therein]{mcintosh99},  although 
the difference is rather large \citep[see also][]{aoki04}.
In addition we also detect the intervening systems at z=1.030 and z=0.559 
\citep[reported also by][]{heidt04},  and the CaII 3934, 3968 \AA \ absorption 
lines from the ISM of our galaxy.
This source exhibits the largest luminosity for the MgII emission line of the 
objects examined here.

\paragraph{1RXS J055806.6$-$383829} The observations of this source performed 
by \citet{giommi89} gave a  featureless optical spectrum. Our spectrum 
clearly shows the host galaxy spectral features (CaII 3934, 3968 \AA, G band 
4305 \AA \ and MgI 5175 \AA \ absorption lines) at z=0.302. We see also several
absorption features from the ISM of our galaxy: DIB at 4726 
\AA \ and 5772 \AA, and the NaI 5892 \AA \ absorption line.
The spectral index found by \citet{giommi89} is $\alpha=$1.8, steeper than our 
result of 1.2.

\paragraph{PKS 0808+019} 
The spectrum obtained by \citet{veron03} shows no 
spectral features. 
A tentative redshift z=0.93 based on a possible detection of MgII 2798 \AA \
was proposed by \citet{jackson02} \citep[see also][]{baldwin81, strittmatter74}.
In our higher S/N  spectrum we detect CIII] 1909 \AA \ and MgII 2798 \AA \ 
emission lines at z=1.148. CaII 3934, 3968 \AA,  NaI 5892 \AA \ and the DIB at 5772 \AA \  
absorption features originated by our galaxy ISM are also observed.

\paragraph{1WGAJ1012.2+063} The optical spectrum obtained by \citet{wolter97} 
failed to detect any feature in this source. 
Our higher S/N spectrum clearly shows two faint emission lines that  we 
identify with MgII 2798 \AA \ and [OII] 3727 \AA \ at z=0.727. We found also 
an intervening absorption line at  4246  \AA \ which we identify as MgII 2798 
\AA \ at z=0.518. This spectrum presents also several absorption features from 
our galaxy ISM: we observe DIB at 4726 
and 5772 \AA, along with CaII 3934, 3968 \AA \ and  NaI 5892 \AA \ atomic lines.

\paragraph{PKS 1250-330} 
Previous spectroscopic observations of this source by \citet{hook03} suggested 
the presence of a weak emission line at 5202 \AA. Our spectroscopy confirms 
with high confidence this broad emission line  which
we identify with MgII 2798 at z=0.856. As in the case of PKS 0306+102, the 
EW of the emission line suggests an intermediate classification between a BL Lac and a QSO.

\paragraph{PKS 1256-229} 
We detect [OII] 3727 \AA ~ and [OIII] 5007 \AA \ 
narrow emission lines at z=0.481. This contrasts with the claim of z=1.365 
proposed by \citet{drinkwater97} without any specification of the observed 
lines. Our visual inspection of the spectrum of \citet{drinkwater97}, 
published on a micro-fiche, does not reveal any significant spectral feature. 
The continuum shape around 3900 \AA \ suggests the presence of the CaII break. 
However, the expected location of the CaII 3934, 3968 \AA \ doublet is 
contaminated by the NaI 5892 \AA ~ 
interstellar absorption. DIB at 4428 \AA \ and 4726 \AA \ are also detected.

\paragraph{PKS 1519-273}
Our VLT spectra yield z=1.297 and confirm the results obtained by 
\citet{heidt04} who detect an emission line  identified as Mg II 2798 \AA \ 
giving z= 1.294. Several absorption features from the ISM of 
our galaxy are observed in this spectrum: DIB at 4428 \AA, 4726 \AA, 4882 \AA
\ and 5772 \AA, along with  CaII 3934, 3968 \AA \ and NaI 5892 \AA \ atomic 
lines.
Although the redshift of this source is based only on a single line, the value 
z=0.07, proposed by \citet{zensus02} (no spectrum published), appears 
inconsistent with our data. 

\paragraph{PKS 2354-021} The spectrum obtained by \citet{hook03} for this 
object is featureless. We clearly detect an emission line 
which we identify as MgII 2798 \AA \ at a redshift z=0.812.  On this spectrum
 we observe also the CaII 3934, 3968 \AA \ absorption feature from our galaxy 
ISM. 

\section{Conclusions}

In this work we have described high S/N spectroscopic observations of 12 BL Lac
objects from which the redshift, in the range 0.2$<$z$<$1.3, has been derived. 
While the targets discussed in this paper do not represent a statistical sample, 
some considerations on the results summarized in table 
\ref{tab:lines} can be proposed.

In three cases we have detected faint (EW are $\sim$1 \AA) 
absorption lines  from the host galaxy. These refer to nearby objects with
0.3$<$z$<$0.5. Only for one of them (1RXS J022716.6+020154) the host galaxy was
also detected by imaging. The others are therefore good candidates for follow 
up imaging studies.

For nine objects the redshift derives from broad and/or narrow emission lines 
in the range 0.2$<$z$<$1.2. For six objects two or more emission 
lines are observed, providing secure determination of the redshift. Only in 
three cases a single broad emission line is detected. The redshift 
for these objects derive from the plausible identification of this line as 
MgII 2798 \AA \ and because of this should be regarded as tentative.  

For seven BL Lac objects we have detected the MgII 2798 \AA \ emission line 
with a  luminosity in the range from  0.5 10$^{42}$erg s$^{-1}$ to 
7.0 10$^{42}$ erg s$^{-1}$, (mean luminosity 1.9 10$^{42}$erg s$^{-1}$). 
These luminosities are about one order of magnitude less than the characteristic 
luminosity of Mg II line in normal quasars \citep{puchnarewicz97}.  
In Figure \ref{fig5} we compare our new measurements with previous 
data  \citep{scarpa97,stickel93} for BL Lacs and blazars over the the continuum
 vs line luminosity plane. 
Our sources cover the lower luminosity region of this plane and conform to the 
behavior of the objects in the class. 
A  correlation between the line and the continuum luminosities is apparent 
over four dex although part of it is induced by the correlation with the 
redshift (partial correlation coefficient L$_{line}$-L$_{cont}$ = 
0.70 after removing the effect due to the correlation with redshift). 
At any given continuum luminosity there is a spread of about 2 dex in line 
luminosity which translates into a difference of about two orders of magnitude 
in the observed EW.
Although the objects appear uniformly distributed in EW, this difference  has 
contributed  to a different classification of the sources (BL Lacs, HPQs, see 
also \citet{scarpa97}). 

\acknowledgments

\textbf{Acknowledgments:} We acknowledge the EC funding under 
contract HPRCN-CT-2002-00321 (ENIGMA network), and the Italian MIUR funding,
 COFIN 2002-027145.

\newpage

\begin{figure}[htbp]
  \centering
  \includegraphics[scale=0.5]{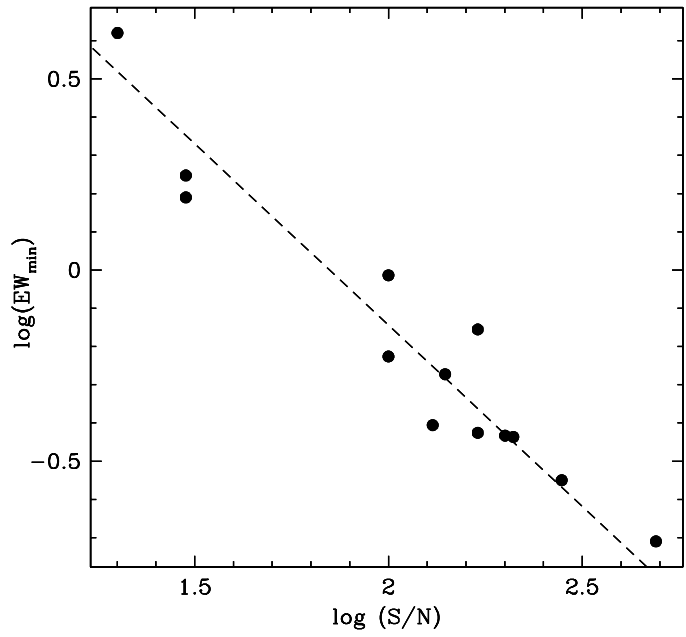}
  \caption{S/N level vs. the minimum detectable equivalent width EW$_{min}$. 
  Points shows the values for the observed spectra. 
  The dashed line shows the EW$_{min} \propto$S/N$^{-1}$ correlation.}
  \label{fig:sn/ew}
\end{figure}

\begin{figure}[htbp]
    \centering
  \resizebox{\hsize}{!}{\includegraphics{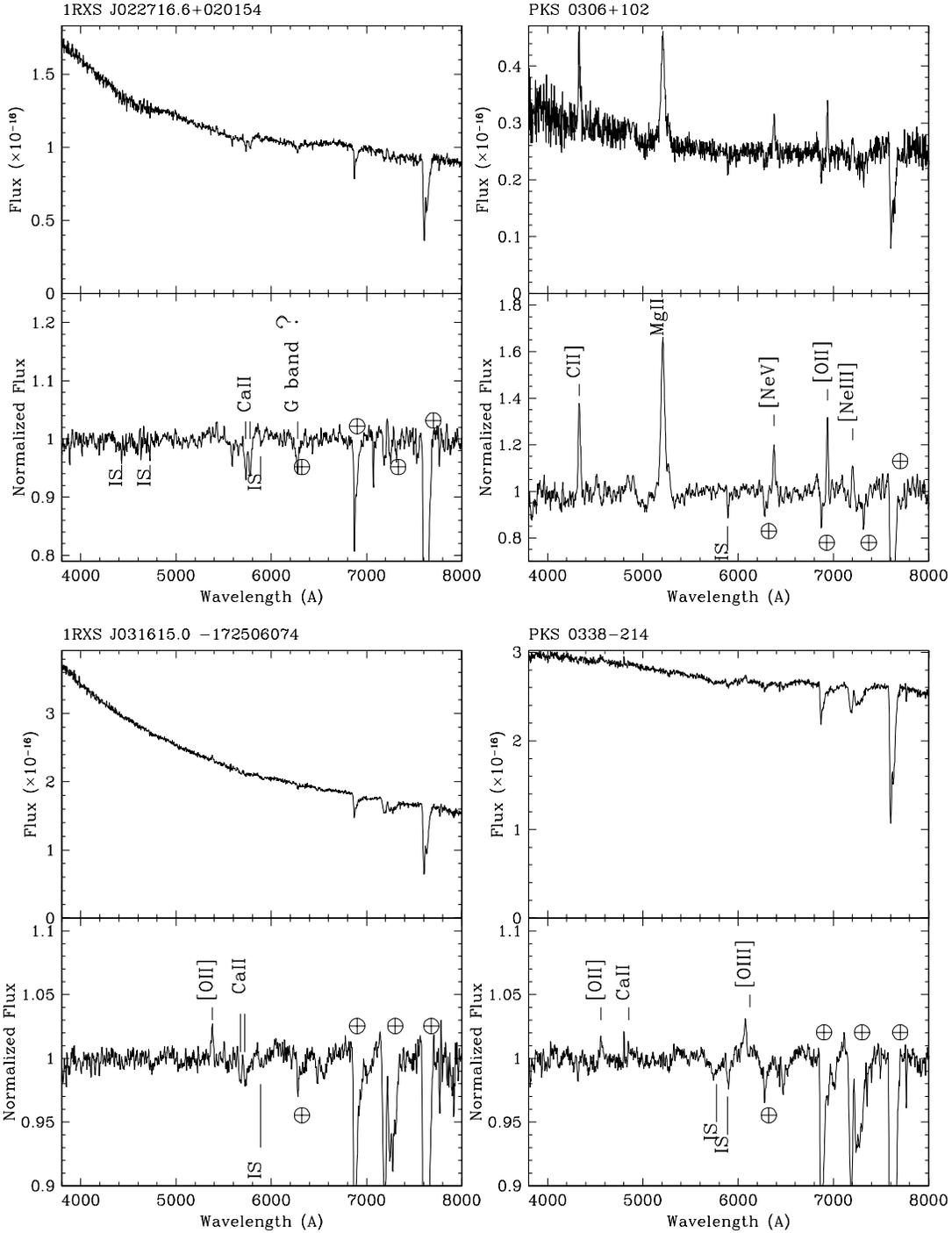}}
  \caption{Spectra of the observed objects. Top panels: flux calibrated and dereddened spectra.
Bottom panels: normalized spectra. Telluric bands are indicated by $\oplus$, spectral lines are marked by the line ID,
intervening MgII absorption systems are reported as "int. MgII",
absorption features in the interstellar medium of our galaxy are labeled by IS.}
\label{fig:spec}
\end{figure}
\addtocounter{figure}{-1}
\begin{figure}[htbp]
  \centering
  \resizebox{\hsize}{!}{\includegraphics{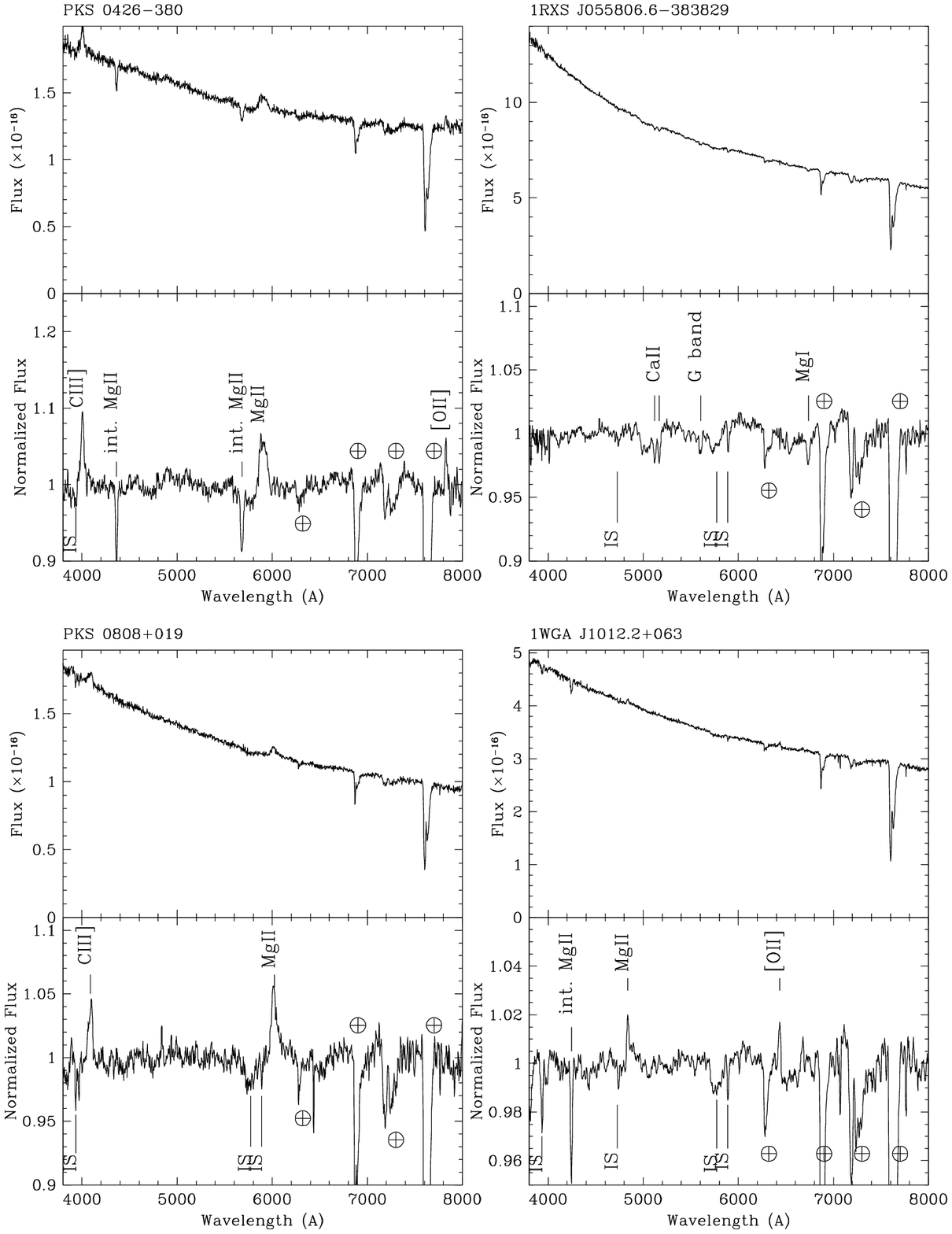}}
  \caption{continued.}

\end{figure}
\addtocounter{figure}{-1}
\begin{figure}[htbp]
    \centering
  \resizebox{\hsize}{!}{\includegraphics{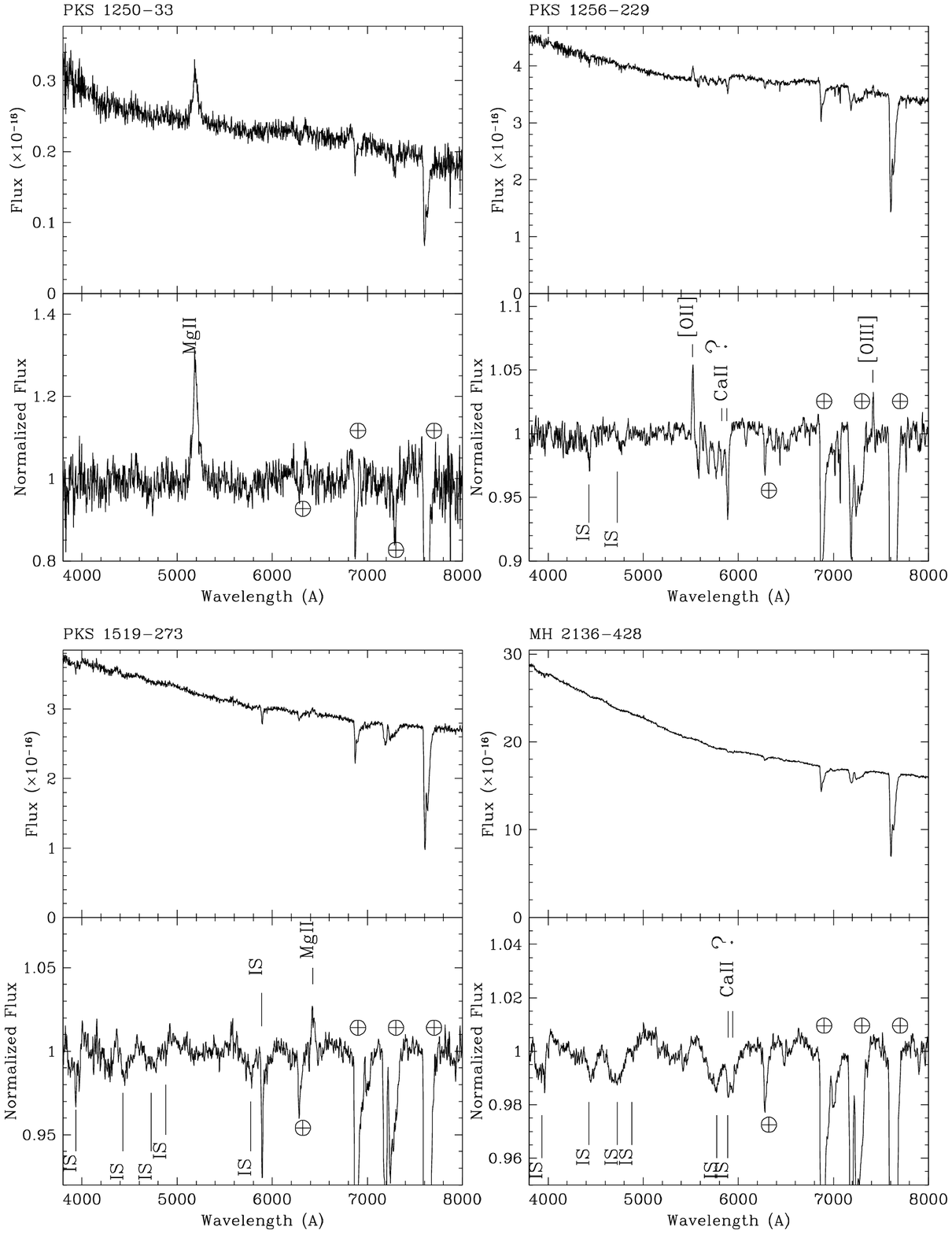}}
  \caption{continued.}

\end{figure}
\begin{figure}[htbp]
  \centering
  \includegraphics[scale=0.6, angle=-90]{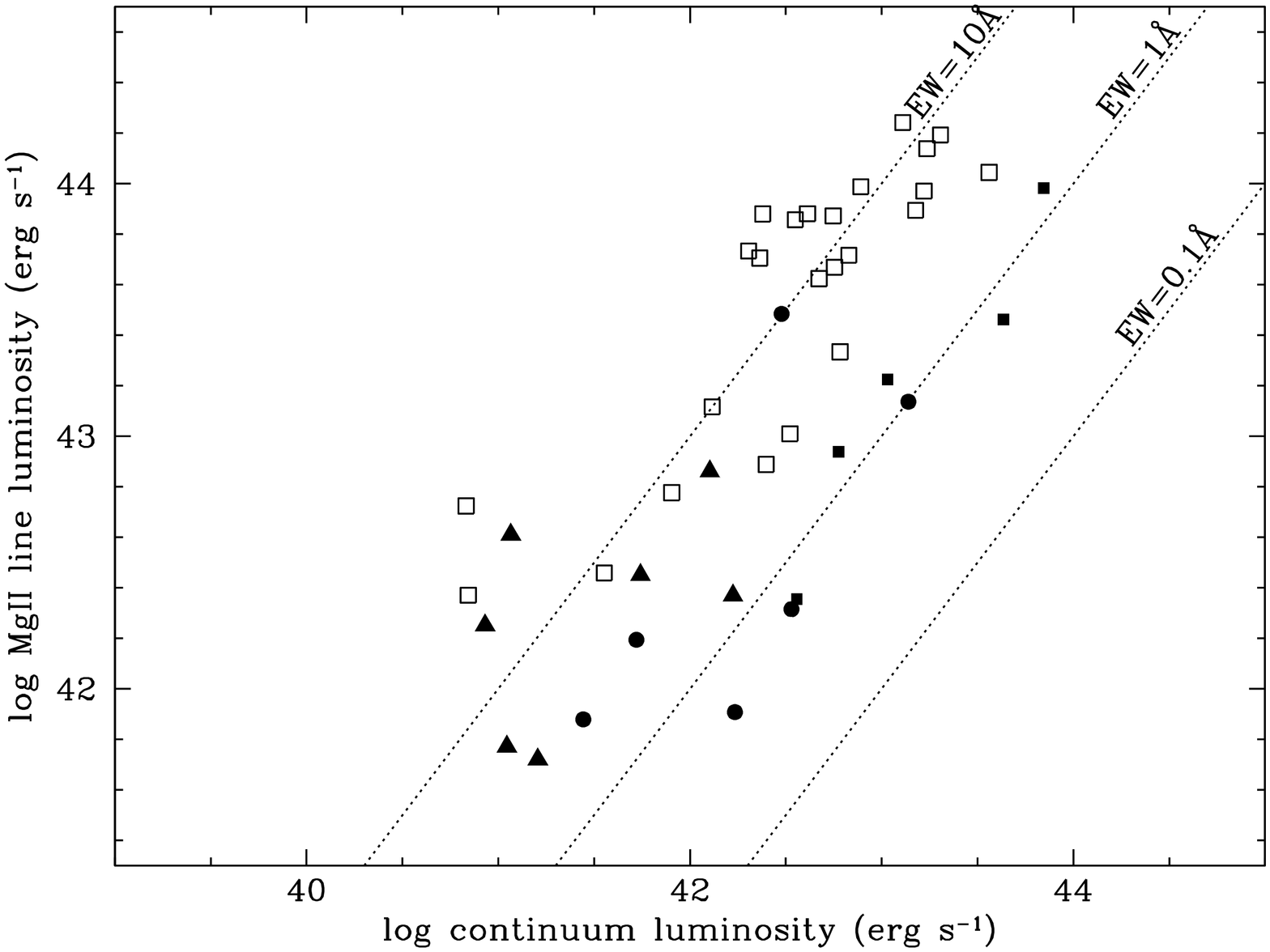}
  \caption{MgII 2798 \AA \ line luminosity versus continuum luminosity.
  Open and filled squares are HPQ and  BL Lacs from \citet{scarpa97},
  filled circles are BL Lacs from \citet{stickel93}, filled triangles are our data. Dotted lines indicate the loci of EW =0.1, 1, 10 \AA.}
  \label{fig5}
\end{figure}

\newpage      

\begin{deluxetable}{lllllllll}
\tabletypesize{\footnotesize}
\tablecaption{Journal of observations\label{tab:list}}
\tablehead{
\colhead{Object}&\colhead{RA}&\colhead{Dec}&\colhead{V}&\colhead{Class}&
\colhead{Date}&\colhead{Exposure}&
\colhead{S/N}&\colhead{Ref.}\\
\colhead{}&\colhead{(J2000)}&\colhead{(J2000)}&\colhead{}&\colhead{}&\colhead{}&
\colhead{time}&\colhead{}&\colhead{}\\
\colhead{(1)}&\colhead{(2)}&\colhead{(3)}&\colhead{(4)}&\colhead{(5)}&\colhead{(6)}&\colhead{(7)}&\colhead{(8)}&\colhead{(9)}\\
}
\startdata
1RXS J022716.6+020154        &02 27 16.6  &+02 01 58.0    &18.8 &H  &24 Dec 03 &2325  &100  &NA96       \\
PKS 0306+102                 &03 09 03.6  &+10 29 12.3    &18.4 &L  &19 Nov 03 &2325  &20   &CO77, VE94 \\
1RXS J031615.0$-$26074       &03 16 15.0  &$-$26 07 56.7  &17.5 &L  &19 Nov 03 &2325  &130  &BA94, BA00 \\
PKS 0338$-$214               &03 40 35.5  &$-$21 19 31.2  &17.1 &L  &19 Nov 03 &2325  &210  &FA94, FA00 \\
PKS 0426$-$380               &04 28 40.4  &$-$37 56 19.6  &19.0 &L  &31 Jul 03 &1800  &100  &ST93, HE04 \\
1RXS J055806.6$-$383829      &05 58 06.2  &$-$38 38 27.0  &17.1 &H  &25 Dec 03 &2325  &280  &KO92       \\
PKS 0808+019                 &08 11 26.7  &+01 46 52.2    &17.2 &L  &25 Dec 03 &2325  &140  &VE93       \\
1WGA J1012.2+063             &10 12 12.2  &+06 31 01.0    &16.8 &L  &30 Dec 03 &2325  &200  &WO97       \\
PKS 1250$-$33                &12 52 58.4  &$-$33 19 59.3  &21.5 &L  &24 Jan 04 &2325  &30   &DR97       \\
PKS 1256$-$229               &12 59 08.5  &$-$23 10 38.7  &16.7 &L  &27 Jan 04 &2325  &170  &DR97       \\
PKS 1519$-$273               &15 22 37.7  &$-$27 30 10.8  &17.7 &L  &30 Apr 03 &2325  &170  &PE98, HE04 \\
PKS 2354$-$021               &23 57 25.1  &$-$01 52 15.3  &21.2 &L  &17 Sep 03 &2325  &30   &HO03       \\
\enddata
\tablecomments{Description of columns: 
(1) Object name; 
(2) Right Ascension (J2000); 
(3) Declination (J2000); 
(4) V magnitude from V\'eron-Cetty \& V\'eron 2003) catalogue; 
(5) Object class (H: High energy peaked BL Lac, L: Low energy peaked BL Lac, 
as defined by \citet{padovani95b}); 
(6) date of observations; 
(7) Exposure time (seconds); 
(8) Signal to Noise; 
(9) Reference to previous optical observations 
CO77: \citet{condon77}; KO92: \citet{kotilainen92}; VE93: 
\citet{veron93}; ST93: \citet{stickel93}; BA94: \citet{bade94}; FA94: \citet{falomo94};
VE94: \citet{veron94}; NA96: \citet{nass96}; DR97: \citet{drinkwater97}; 
WO97: \citet{wolter97}; PE98: \citet{perlman98}; BA00:\citet{bauer00}; 
FA00: \citet{falomo00}; HO03: \citet{hook03}; HE04: \citet{heidt04}.}
\end{deluxetable}

\begin{deluxetable}{lccclcccccc}
\tabletypesize{\footnotesize}
\tablecaption{Spectral lines parameters.\label{tab:lines}}
\tablehead{
\colhead{Object name}&\colhead{z$_{avg}$}&\colhead{$\alpha$}&\colhead{V}&\colhead{Line ID}&\colhead{$\lambda$}&\colhead{z}&\colhead{Type}&\colhead{FWHM}&\colhead{EW}&\colhead{Line}\\
\colhead{}&\colhead{}&\colhead{}&\colhead{}&\colhead{}&\colhead{}&\colhead{}&\colhead{}&\colhead{}&\colhead{}&\colhead{luminosity}\\
\colhead{(1)}&\colhead{(2)}&\colhead{(3)}&\colhead{(4)}&\colhead{(5)}&\colhead{(6)}&\colhead{(7)}&\colhead{(8)}&\colhead{(9)}&\colhead{(10)}&\colhead{(11)}\\}
\startdata
                &       &      &       &          &      &       &   &       &        &     \\
1RXSJ022716.6+020154&0.457  &0.83  &18.9   &      &      &       &   &       &        &     \\
                &       &      &       &Ca II     &5731  &0.457  &g  &1000   &$+$1.6  &     \\
                &       &      &       &Ca II     &5779  &0.456  &g  &1000   &$+$1.7  &     \\
                &       &      &       &G band    &6277  &0.458  &g  &900    &$+$1.3  &     \\
PKS 0306+102    &0.862  &0.47  &21.2   &          &      &       &   &       &        &     \\
                &       &      &       &C II]     &4332  &0.862  &e  &1800   &$-$10   &15.6\\
                &       &      &       &Mg II     &5210  &0.862  &e  &2400   &$-$35   &40.7\\
                &       &      &       &[Ne V]    &6376  &0.862  &e  &600    &$-$1    &4.9 \\
                &       &      &       &[O II]    &6938  &0.862  &e  &500    &$-$5    &5.0 \\
                &       &      &       &[Ne III]  &7202  &0.862  &e  &800    &$-$6    &1.9 \\
1RXSJ031615.0$-$260748&0.443  &1.20  &18.1   &    &      &       &   &       &        &     \\
                &       &      &       &[O II]    &5377  &0.443  &e  &1500   &$-$0.6  &1.2 \\
                &       &      &       &CaII      &5678  &0.443  &g  &2200   &$+$0.6  &     \\
                &       &      &       &CaII      &5724  &0.442  &g  &1500   &$+$0.6  &     \\
                &       &      &       &G band    &6213  &0.443  &g  &1200   &$+$0.6  &     \\
PKS 0338$-$214  &0.223  &0.27  &17.9   &          &      &       &   &       &     &        \\
                &       &      &       &[O II]    &4560  &0.223  &e  &4700   &$-$0.6  &0.63 \\
                &       &      &       &[O III]   &6074  &0.224  &e  &2400   &$-$1.8  &0.52 \\
PKS 0426$-$380  &1.105  &0.70  &18.6   &          &      &       &   &       &        &     \\
                &       &      &       &C III]    &4006  &1.098  &e  &3000   &$-$3.6  &47.7\\
                &       &      &       &Mg II     &5908  &1.112  &e  &4700   &$-$5.7  &71.8\\
                &       &      &       &[O II]    &7826  &1.099  &e  &1000   &$-$1.2  &16.2\\
                &       &      &       &MgII      &4362  &0.559  &a  &1200   &$+$2.2  &     \\
                &       &      &       &MgII      &5681  &1.030  &a  &1500   &$+$2.3  &     \\
1RXSJ055806.6$-$383829&0.302  &1.21  &16.8   &    &      &       &   &       &        &     \\
                &       &      &       &Ca II     &5120  &0.301  &g  &1400   &$+$0.9  &     \\
                &       &      &       &Ca II     &5169  &0.302  &g  &1400   &$+$0.8  &     \\
                &       &      &       &G band    &5605  &0.302  &g  &1800   &$+$0.7  &     \\
                &       &      &       &Mg I      &6735  &0.302  &g  &1600   &$+$0.8  &     \\
PKS 0808+019    &1.148  &0.93  &18.4   &          &      &       &   &       &        &     \\
                &       &      &       &C III]    &4099  &1.147  &e  &4200   &$-$2.8  &31.0\\
                &       &      &       &Mg II     &6014  &1.149  &e  &5400   &$-$5.1  &28.34\\
1WGAJ1012.2+063 &0.727  &0.80  &17.6   &          &      &       &   &       &        &     \\
                &       &      &       &Mg II     &4838  &0.729  &e  &2000   &$-$0.6  &5.3 \\
                &       &      &       &[O II]    &6434  &0.726  &e  &2100   &$-$0.6  &4.9 \\
                &       &      &       &MgII      &4246  &0.518  &a  &1400   &$+$1.2  &     \\
PKS 1250$-$33   &0.856* &0.48  &20.1   &          &      &       &   &       &        &     \\
                &       &      &       &Mg II     &5192  &0.856  &e  &4200   &$-$20.8 &17.8\\
PKS 1256$-$229  &0.481  &0.31  &17.9   &          &      &       &   &       &        &     \\
                &       &      &       &[O II]    &5521  &0.481  &e  &1200   &$-$1.5  &4.6 \\
                &       &      &       &[O III]   &7417  &0.481  &e  &500    &$-$0.7  &1.5 \\
PKS 1519$-$273  &1.297* &0.51  &17.8   &          &      &       &   &       &        &     \\
                &       &      &       &Mg II     &6427  &1.297  &e  &1900   &$-$1.4  &23.3\\
PKS 2354$-$021  &0.812* &0.67  &20.4   &          &      &       &   &       &        &     \\
                &       &      &       &Mg II     &5071  &0.812  &e  &6100   &$-$5.3  &5.8 \\
\enddata 
\tablecomments{Description of columns: 
(1) Name of the source; 
(2) average of redshift from the single lines; 
(3) spectral $\alpha$ index of the continuum, defined by 
F$_{\lambda}\propto\lambda^{-\alpha}$; 
(4) V magnitude estimated from observed spectra; 
(5) line identification; 
(6) observed wavelength of line center(\AA); 
(7) redshift of the line; 
(8) type of the line (\textbf{e}: emission line; 
\textbf{g}: absorption line from the host galaxy;
\textbf{a}: absorption line from intervening systems); 
(9) FWHM of the line (km s$^{-1}$); 
(10) EW of the line (\AA); 
(11) emission line luminosity (10$^{41}$erg s$^{-1}$).
*: tentative redshift estimate, obtained assuming that the only detected emission feature is 
MgII 2798 \AA.}
\end{deluxetable}


\newpage

\begin{thebibliography}

\bibitem[Aoki, Kawaguchi, \& Ohta(2004)]{aoki04} Aoki, K., Kawaguchi, T., 
Ohta, K., astro-ph/0409546

\bibitem[Appenzeller et al.(1998)]{fors} Appenzeller et al., Messenger 94,
 1, 1998

\bibitem[Bade, Fink, \& Engels(1994)]{bade94} Bade, N., Fink, 
H.~H., \& Engels, D.\ 1994, \aap, 286, 381 

\bibitem[Baldwin, Wampler, \& Burbidge(1981)]{baldwin81} 
Baldwin, J.~A., Wampler, E.~J., \& Burbidge, E.~M.\ 1981, \apj, 243, 76 

\bibitem[Bauer, Condon, Thuan, \& Broderick(2000)]{bauer00} 
Bauer, F.~E., Condon, J.~J., Thuan, T.~X., \& Broderick, J.~J.\ 2000, 
\apjs, 129, 547 

\bibitem[Carangelo et al.(2003)]{carangelo03} Carangelo, N., 
Falomo, R., Kotilainen, J., Treves, A., \& Ulrich, M.-H.\ 2003, \aap, 412, 
651 

\bibitem[Cardelli, Clayton, \& Mathis(1989)]{cardelli89} Cardelli, 
J.~A., Clayton, G.~C., \& Mathis, J.~S.\ 1989, \apj, 345, 245 

\bibitem[Condon, Hicks, \& Jauncey(1977)]{condon77} Condon, 
J.~J., Hicks, P.~D., \& Jauncey, D.~L.\ 1977, \aj, 82, 692 

\bibitem[Drinkwater et al.(1997)]{drinkwater97} Drinkwater, M.~J., 
et al.\ 1997, \mnras, 284, 85 

\bibitem[Falomo \& Ulrich(2000)]{falomo00} Falomo, R.~\& Ulrich, 
M.-H.\ 2000, \aap, 357, 91 

\bibitem[Falomo, Scarpa, \& Bersanelli(1994)]{falomo94} Falomo, 
R., Scarpa, R., \& Bersanelli, M.\ 1994, \apjs, 93, 125 

\bibitem[Giommi et al.(1989)]{giommi89} Giommi, P., Beuermann, 
K., Barr, P., Schwope, A., Tagliaferri, G., \& Thomas, H.~C.\ 1989, \mnras, 
236, 375 

\bibitem[Giommi, Ansari, \& Micol(1995)]{giommi95} Giommi, P., 
Ansari, S.~G., \& Micol, A.\ 1995, \aaps, 109, 267 

\bibitem[Heidt et al.(2004)]{heidt04} Heidt, J., Tr{\" o}ller, 
M., Nilsson, K., J{\" a}ger, K., Takalo, L., Rekola, R., \& Sillanp{\" 
a}{\" a}, A.\ 2004, \aap, 418, 813 

\bibitem[Hook et al.(2003)]{hook03} Hook, I.~M., Shaver, 
P.~A., Jackson, C.~A., Wall, J.~V., \& Kellermann, K.~I.\ 2003, \aap, 399, 
469

\bibitem[Jackson et al.(2002)]{jackson02} Jackson, C.~A., Wall, 
J.~V., Shaver, P.~A., Kellermann, K.~I., Hook, I.~M., \& Hawkins, M.~R.~S.\ 
2002, \aap, 386, 97 

\bibitem[Jauncey, et al.(1982)]{jauncey82} 
Jauncey, D.~L., Batty, M.~J., Gulkis, S., \& Savage, A.\ 1982, \aj, 87, 763 

\bibitem[Kinney et al.(1996)]{kinney96} Kinney, A.~L., Calzetti, 
D., Bohlin, R.~C., McQuade, K., Storchi-Bergmann, T., \& Schmitt, H.~R.\ 
1996, \apj, 467, 38 

\bibitem[Kotilainen et al.(1992)]{kotilainen92} Kotilainen, J.~K., 
Ward, M.~J., Boisson, C., Depoy, D.~L., Smith, M.~G., \& Bryant, L.~R.\ 
1992, \mnras, 256, 125 

\bibitem[Landt et al.(2001)]{landt01} Landt, H., Padovani, P., Perlman, E.~S.,
  Giommi, P., Bignall, H. \& Tzioumis, A.\ 2001, \mnras, 323, 757

\bibitem[Ledden \& O'Dell(1985)]{ledden85} Ledden, J.~E.~\& 
O'Dell, S.~L.\ 1985, \apj, 298, 630 

\bibitem[Londish et al.(2002)]{londish02} Londish, D., et al.\ 
2002, \mnras, 334, 941 

\bibitem[Marcha et al.(1996)]{marcha96} 
Marcha, M.~J.~M., Browne, I.~W.~A., Impey, C.~D., \& Smith, P.~S.\ 1996, 
\mnras, 281, 425 

\bibitem[McIntosh et al.(1999)]{mcintosh99} 
McIntosh, D.~H., Rix, H.-W., Rieke, M.~J., \& Foltz, C.~B.\ 1999, \apjl, 
517, L73 

\bibitem[Nass et al.(1996)]{nass96} Nass, P., Bade, N., 
Kollgaard, R.~I., Laurent-Muehleisen, S.~A., Reimers, D., \& Voges, W.\ 
1996, \aap, 309, 419 

\bibitem[Nilsson et al.(2003)]{nilsson03} Nilsson, K., Pursimo, T., Heidt, J.,
  Takalo, L.~O., Sillanp{\"a}{\"a}, A., \& Brinkmann, W.\ 2003 \aap, 400, 95

\bibitem[Oke(1990)]{oke90} Oke, J.~B.\ 1990, \aj, 99, 1621

\bibitem[Padovani \& Giommi(1995)a]{padovani95a} Padovani, P.~\& 
Giommi, P.\ 1995, \mnras, 277, 1477 

\bibitem[Padovani \& Giommi(1995)b]{padovani95b} Padovani, P.~\& 
Giommi, P.\ 1995, \apj, 444, 567 

\bibitem[Perlman et al.(1998)]{perlman98} Perlman, E.~S., 
Padovani, P., Giommi, P., Sambruna, R., Jones, L.~R., Tzioumis, A., \& 
Reynolds, J.\ 1998, \aj, 115, 1253 

\bibitem[Puchnarewicz et al.(1997)]{puchnarewicz97} Puchnarewicz, 
E.~M., et al.\ 1997, \mnras, 291, 177 

\bibitem[Rector \& Stocke(2001)]{rector01} Rector, T.~A.~\& 
Stocke, J.~T.\ 2001, \aj, 122, 565 

\bibitem[Scarpa \& Falomo(1997)]{scarpa97} Scarpa, R.~\& Falomo, 
R.\ 1997, \aap, 325, 109 

\bibitem[Scarpa et al.(2000)]{scarpa00} Scarpa, R., Urry, C.~M., 
Falomo, R., Pesce, J.~E., \& Treves, A.\ 2000, \apj, 532, 740 

\bibitem[Schlegel, Finkbeiner, \& Davis(1998)]{schlegel98} 
Schlegel, D.~J., Finkbeiner, D.~P., \& Davis, M.\ 1998, \apj, 500, 525 

\bibitem[Stickel, Fried, \& Kuehr(1993)]{stickel93} Stickel, M., 
Fried, J.~W., \& Kuehr, H.\ 1993, \aaps, 98, 393 


\bibitem[Strittmatter, Carswell, \& Gilbert(1974)]{strittmatter74} 
Strittmatter, P.~A., Carswell, R.~F., \& Gilbert, G.\ 1974, \apj, 190, 509 

\bibitem[Tody(1993)]{tody93} Tody, D.\ 1993, ASP Conf.~Ser.~ 
52: Astronomical Data Analysis Software and Systems II, 2, 173 

\bibitem[Tody(1986)]{tody86} Tody, D.\ 1986, \procspie, 627, 
733 

\bibitem[Urry et al.(2000)]{urry00} Urry, C.~M., Scarpa, R., 
O'Dowd, M., Falomo, R., Pesce, J.~E., \& Treves, A.\ 2000, \apj, 532, 816 

\bibitem[Valdes(1992)]{valdes92} Valdes, F.\ 1992, ASP 
Conf.~Ser.~ 25: Astronomical Data Analysis Software and Systems I, 1, 417 

\bibitem[V{\'e}ron(1994)]{veron94} V{\'e}ron, P.\ 1994, \aap, 283, 802 

\bibitem[V{\' e}ron-Cetty \& V{\' e}ron(2003)]{veron03} V{\' e}ron-Cetty, 
M.-P.~\& V{\' e}ron, P.\ 2003, \aap, 412, 399 

\bibitem[V{\'e}ron-Cetty \& V{\'e}ron(1993)]{veron93} V{\'e}ron-Cetty, 
M.-P.~\& V{\'e}ron, P.\ 1993, \aaps, 100, 521 

\bibitem[White, Giommi, \& Angelini(2000)]{white00} White, 
N.~E., Giommi, P., \& Angelini, L.\ 2000, VizieR Online Data Catalog, 9031, 
0 
\bibitem[Wolter et al.(1997)]{wolter97} Wolter, A., et al.\ 
1997, \mnras, 284, 225 

\bibitem[Wright, et al.(1977)]{wright77} 
Wright, A.~E., Jauncey, D.~L., Peterson, B.~A., \& Condon, J.~J.\ 1977, 
\apjl, 211, L115 

\bibitem[Zensus et al.(2002)]{zensus02} Zensus, J.~A., Ros, E., 
Kellermann, K.~I., Cohen, M.~H., Vermeulen, R.~C., \& Kadler, M.\ 2002, 
\aj, 124, 662 

\end{thebibliography}
\end{document}